# SINGULARITY AVOIDANCE IN NUMERICAL BLACK HOLE SPACETIMES[*]

**PETER ANNINOS**[1], **GREG DAUES**[2], **JOAN MASSÓ**[1,3], **EDWARD SEIDEL**[1], **WAI-MO SUEN**[2]

[1] *National Center for Supercomputing Applications*
*605 E. Springfield Ave., Champaign, Illinois 61820*

[2] *McDonnell Center for the Space Sciences, Department of Physics*
*Washington University, St. Louis, Missouri, 63130*

[3] *Departament de Física, Universitat de les Illes Balears,*
*E-07071 Palma de Mallorca, Spain*

ABSTRACT

Spacetime singularities in numerical relativity can be avoided by excising a region of the computational domain from inside the apparent horizon. We report on results of such a scheme that is based on using (*i*) a horizon locking coordinate which locks the coordinate system to the geometry, and (*ii*) a finite differencing scheme which respects the causal structure of the spacetime. With this technique a black hole can be evolved accurately well beyond $t = 1000M$, where $M$ is the black hole mass.

## 1. Background

As black holes are accompanied by singularities, their presence in numerical spacetimes leads to extreme dynamic ranges in length and time, making it difficult to maintain accuracy and stability for long periods of time. The traditional way to deal with these problems has been to take advantage of the coordinate degrees of freedom inherent in the Einstein equations to avoid evolving the extreme curvature regions. These so-called singularity avoiding slicing conditions wrap up around the singular region so that a large fraction of the spacetime outside the singular region can be evolved. However, these conditions by themselves do not completely solve the problem; they merely serve to delay the breakdown of the numerical evolution. In the vicinity of the singularity, these slicings inevitably contain a region of abrupt change near the horizon and a region in which the constant time slices dip back deep into the past in some sense. Numerical simulations will eventually crash due to these pathological properties of the slicing. As these problems are even more severe in 3D, where much longer evolutions will be required to study important problems like the coalescence of two black holes, it is essential to investigate alternative methods to handle singularities and black holes in numerical relativity.

## 2. Horizon Boundary Condition

Cosmic censorship suggests that in physical situations, singularities are hidden inside black hole horizons. Because the region of spacetime inside the horizon cannot causally affect the region of interest outside the horizon, one is tempted to cut away

---

the interior region containing the singularity and evolve only the singularity-free region outside. To an outside observer no information will be lost since the region cut away is unobservable. The procedure of cutting away the singular region will drastically reduce the dynamic range, making it easier to maintain accuracy and stability. With the singularity removed from the numerical spacetime, there is in principle no physical reason why black hole codes cannot be made to run indefinitely to cover all spacetime region outside the horizon of interest without crashing.

Although the desirability of a horizon boundary condition has been raised many times in the literature,[1-4] it has proved to be difficult to implement such a scheme in a dynamical evolution. Recently we demonstrated that a horizon boundary condition can be realized.[5,6] There are two basic ideas behind our implementation of the inner boundary condition. First, we use a "horizon locking coordinate" which locks the spatial coordinates to the spatial geometry and causal structure. This amounts to using a shift vector that locks the horizon in place near a particular coordinate location, and also keeps other coordinate lines from drifting towards the hole. We have investigated several types of shift conditions, including a "distance freezing" shift that freezes the proper distance to the horizon, an "expansion freezing" shift that freezes the rate of expansion of outgoing null rays, an "area freezing" shift that freezes the area of radial shells, and the minimal distortion shift[7] that minimizes the global distortion in the 3-metric. Some of these shifts have the advantage that they can be generalized more easily to geometries and coordinate systems other than the spherical one. The basic message is that the idea of a horizon locking coordinate is robust enough for many different implementations, with some implementations likely extendible to the general 3D case.

The second ingredient is using a finite differencing scheme that respects the causal structure of the spacetime, which essentially means that spatial derivatives are computed at the "center of the causal past" of the point being updated. Such a differencing scheme is not only essential for the stability of codes using large shift vectors as in those with "horizon locking coordinates", but also eliminates the need of explicitly imposing boundary conditions on the horizon. Since the horizon is a one-way membrane, quantities on the horizon can be affected only by quantities outside but not inside the horizon. Hence, in a finite differencing scheme which respects the causal structure, all quantities on the horizon can be updated solely in terms of known quantities residing on or outside the horizon, and there is no need to impose boundary conditions to account for information not covered by the numerical evolution.

It is informative to compare results in using horizon locking schemes against those obtained with traditional singularity avoiding slicing methods. In Figure 1 we show the deviation of the apparent horizon mass for a spherically symmetric black hole from the analytic value of 2. The solid line is the deviation for the "distance freezing" shift vector[6] and the dashed line is computed from a code using maximal slicing with zero shift.[8] By $t \sim 100M$, where $M$ is the black hole mass, the zero shift case has accumulated a 100% error in the mass and the code crashes shortly thereafter. The horizon locking shift case is accurate to $\sim 4\%$ at $t = 1000M$.

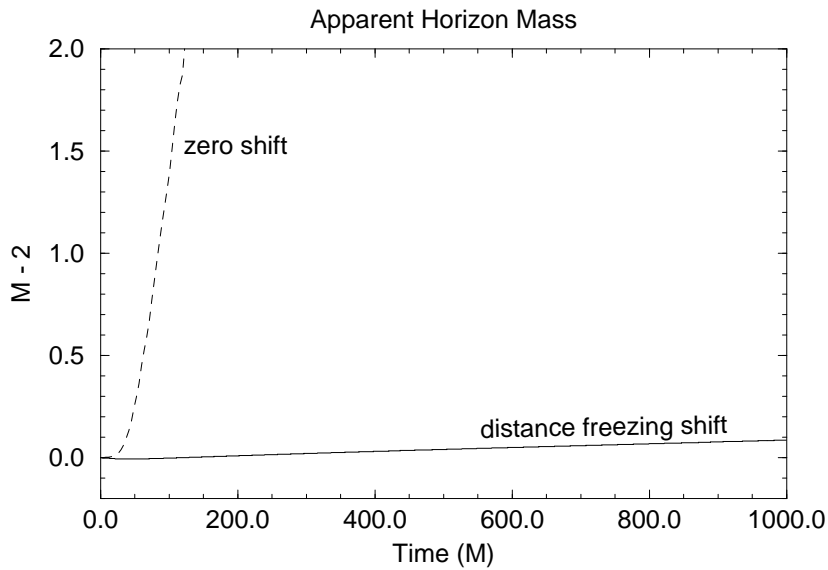

Figure 1: The deviation in the horizon mass for a spherically symmetric black hole is shown as a function of time. We compare the distance freezing shift case to a maximally sliced evolution with zero shift. The zero shift evolution crashes shortly after $t = 100M$.

## Acknowledgements

We are happy to acknowledge helpful discussions with Andrew Abrahams, David Bernstein, Matt Choptuik, David Hobill, Ian Redmount, Larry Smarr, Jim Stone, Kip Thorne, Lou Wicker, and Clifford Will. J.M. acknowledges a Fellowship (P.F.P.I.) from Ministerio de Educación y Ciencia of Spain. This research is supported by the NCSA, the Pittsburgh Supercomputing Center, and NSF grants Nos. PHY91-16682, PHY94-04788, PHY94-07882 and PHY/ASC93-18152 (arpa supplemented).